\documentstyle[12pt]{article}

\newcommand{\be}{\begin{equation}}
\newcommand{\ee}{\end{equation}}
\newcommand{\ba}{\begin{eqnarray*}}
\newcommand{\ea}{\end{eqnarray*}}
\newcommand{\bal}{\begin{eqnarray}}
\newcommand{\eal}{\end{eqnarray}}

\title{How to reconcile Market Efficiency and Technical 
Analysis}

\author{Alexandra Ilinskaia$^1$\  and\  
Kirill Ilinski$^2$\thanks{E-mail: kni@th.ph.bham.ac.uk}
\\ [1cm]
{\small\it $^1$ Quantitative Analysis,
Australia and New Zealand Investment Bank, 
}\\
{\small\it London office, United Kingdom}
\\ [0.5cm]
{\small\it $^2$ School of Physics and Astronomy,
University of Birmingham,} \\
{\small\it Edgbaston B15 2TT, Birmingham, United Kingdom}
}

\date{  }

\begin{document}
\maketitle

\begin{abstract}
Weak form of the Efficiency Market Hypothesis  (EMH) excludes
predictions of future market movements from historical data and
makes the technical analysis (TA) out of law. However the technical analysis
is widely used by traders and speculators who steadely refuse to consider
the market as a "fair game" and survive with such believe. 
In the paper we make a conjecture that TA and EMH correspond to different time
regimes and show how both technical analysis predictions for short times 
and realistic statistical data for larger times can be obtained in a simple 
single stock model of Gauge Theory of Arbitrage. 
\end{abstract}


\section{Introduction}
For many decades people who deal with securites are divided 
into two groups. The first group "feels" the market, listens
how the market "breathes" and treat the market as alive being
\cite{Met}. To do this they analyze historical data for prices and
volumes, draw patterns and construct indicators, i.e. make use of 
machinery of the technical analysis (TA)~\cite{Chande,Achelis}. 
They are technicians.
If somebody asks them whether the price is a random process the answer 
will be emotional and strongly negative. No one trader would agree 
that his job is equivalent to throwing a dice. 
In the same time, the second group, roughly speaking, assumes that the
price is a random process, the game is fair and the market is efficient.
Indeed, even weak form of the Efficient Market Hypothesis (EMH) 
sais~\cite{Blake} that the all relevant information came from historical data 
is encoded in the current price and, hence, the only ingredient which
is able to influence the future prices is a new information. The information
is unpredictable and random. This excludes predictions of future market 
movements from historical data, i.e. makes the technical analysis out of law. 
EMH lais in a basis of the financial analysis~\cite{Blake,Lumby,Duffie} with 
many outcomes such as modern portfolio theory~\cite{portfolio} and
derivative pricing~\cite{Hull}.

The conflict lasts for years. Many efforts have been made to check EMH.
This leaded to believe that "the evidence for weak-form market efficiency
is very strong" and that "technical analysts are deluding themselves about 
their ability to predict future price movements"~\cite{Blake}. 
In the same time, there is a number of statistical estimations of
TA prediction accuracy (see, for example~\cite{Morris,II}) which
excludes pure random process. Shortly, TA more often is right than wrong
but still it is not a "Holy Grail"~\cite{Pring}.
  
Summarizing, traders and speculators mainly stick with technical analysis while
fund managers and quantitative analysts base their strategies on EMH. There
difference between players suggests that the conflict is due to different
time horizons. From this point of view, the TA predictions exists due to
an internal deterministic dynamics which brings the market to equilibrium.
The market has a long enough memory~\cite{Sornette1} and the dynamics is not 
so fast as EMH assumes. We show in the paper that, in fact, TA and EMH can
be observed in the same model for a market relaxation and the regime depends
only on a time scale. More precise, we show that TA indicators and the 
corresponding predictions do exist for short times while for large times
the model produces EMH state with realistic statistical data.

In a certain sense, the situation with EMH and TA 
slightly resembles a conflict between
Schr\"odinger and Laplacian determinism in the beginning of the century in 
physics. Let us remind that the Laplacian (or classical) picture implies the
possibility to determine a future state of a dynamical system precisely subject
to the condition that the current state is completely known. 
The Schr\"odinger's determinism implies that only the probability of 
the future states of the system can be predicted. The conflict is due to the
fact that all we can see around us is deterministic 
(exactly as extreme apologist
of TA may say) but the theory predicts vital influence of quantum fluctuations 
for long enough times (for example, a car can tunel through a wall if we 
wait long enough). However, for small enough times any quantum system 
behaves as classical. Following this line we show in the framework of
Gauge Theory of Arbitrage that as time goes the
deterministic TA regime is subtituted by EMH regime in course of a 
relaxation equilibration process.

Recently, in papers~\cite{GTA,KI} the Gauge Theory of Arbitrage (GTA)
was introduced to describe relaxation processes in financial market.
The approach uses methods of
quantum field theory and based on the observation that the discounting 
procedure can be considered as a parallel transport in a complicated space
and curvature of this space is connected with virtual arbitrage opportunities.
This allows us to map the theory of capital market onto the theory of 
quantum system of particles with positive (securities) and negative ("debts")
charges which interact with each other through electromagnetic field
(gauge field of the arbitrage). In the case of a local virtual
mispricing money flows in the region of configuration
space (money poor in the profitable security) while "debts" try
to escape from the region. Entering positive charges and leaving
negative ones screen up the profitable fluctuation and restore the
equilibrium in the region where there is no mispricing any
more, i.e. speculators washed out the profitable opportunity. It is important 
to note that in simplest approximation of absence of money flows 
GTA  is equivalent to the assumption about the log-normal walks of assets 
prices. More realistic statistical characteristics of price series appear
if the money flows are taken into account~\cite{ISfin2}. We discuss GTA in 
more details in section 4.

The paper is organized as follows. In the next section we main issues
of EMH and TA to make the consideration self-contained. In section 3
we give motivation of GTA consideration and introduce the GTA model 
following~\cite{GTA,KI,ISfin2}. We show how the model leads to
TA behaviour in the short time limit and demonstrates a quantitative agreement
with market statistical data for large enough times. The paper is concluded with 
final remarks.

\section{Technical Analysis vs Market Efficiency}

\subsection{Technical Analysis}
The technical analysis (TA) in general can be defined as a set of methods
for predictions of future prices which is based on``mathematical" rather than
economical calculations. It was founded for pure utilitarian purposes, i.e.
to get a profit from speculations playing with stocks and, later, with
futures. In the same measure as the fundamental analysis is a job for 
economists, the technical analysis is a field for engineers. 
It does not mean, however, that there is not place for a fundamental 
economic information in TA. In paper~\cite{II} we showed how to incorporate
fundamental informations on a correlation of companies share prices
into TA to construct new correlative indicators on the example of two 
principle energy supply companies in Russia. We believe that the background for
a general multiasset consideration is given by Mantegna's ultrametric 
trees~\cite{Mantegna} as it was constructed for stocks in DJIA and 
S\& P500 index.

Keeping aside patterns of classical technical 
analysis, Elliott wave theory~\cite{Elliott} and 
Japanise candlesticks~\cite{Morris} where a recognition of a pattern is quite
subjective, it is possible to say that TA consists of a set of simple
indicators calculated from previous volumes and prices. The indicators produce
``buy" and ``sell" signals and can be verified on historical data (this does not
implies, however, that the indicators will work in future). It is clear that
there is a huge number of combinations of prices and volumes which can be
potential indicators but only few of them survive after real data tests.

Mathematically TA is very simple and cannot be compared with complicated
financial mathematics involved in portfolio theory, pricing and hedging theory.
This causes a certain disbelief in TA. Another source of doubts is a possibility
to construct a trading plan, i.e. an algorithm which will substitute a trader.
Indeed, the market consists of human beings, reflects human psychology and
cannot be put in a finite set of simple equations. Supporters of TA  argue
that the basic principles of economic theories are not very 
complicated as well though they developed to describe the same market.
Concerning the trading plan, there is an opinion that the plan is useful
to save financial and emotional potential of traders in uncertain and
fastly changing informational enviroment~\cite{Open}.
In a sense it is analogues to walking in a forest when somebody lost his way 
-- it is better to go straight ahead instead of making loops after loops
as a drunk sailor.

Putting short, TA being a contraversial subject, is popular among investors and
widely used by professional traders. It is not risk-free. Clear signals do not
come very often and it is hard to collect statistics and make money of it. 
The situation on a market changes with the time and it is important to adjust
TA toolkit to it. This fact also smears the statistics. However, TA is more 
often right than wrong and there are certain market patterns.

Let us now to describe a few TA indicators to illustrate that was said above.
First we consider William's \% R indicator $W_n$ for $n$ periods (e.g. days)
which constructed from previous prices only.
It is given by the formula:
$$
W_n = -100\cdot \frac{H_n - C}{H_n - L_n}
$$
where $C$ is a last closing price, $H_n$ and $L_n$ are highest and lowest
prices for last $n$ days. The indicator oscillates between in the range
between -100 and 0. Undervalued zone lays in range -100 and -80(-70), overvalued
zone spans from -20 to 0. The "buy" ("sell") signal comes when the indicator
leaves the undervalued (overvalued) zone. The method is very simple and is 
easy to use. 

Another example of TA toolkit are 
Positive Volume Index ($PVI$) and Negative Volume Index 
($NVI$)~\cite{Achelis} which we use in the paper.
They are defined by the following rules:
\begin{equation}
PVI_n = PVI_{n-1} (1+ \theta (V_n - V_{n-1})\cdot r_n) \ ,
\label{PVI}
\end{equation}
\begin{equation}
NVI_n = NVI_{n-1} (1+ \theta (V_{n-1} - V_n)\cdot r_n) \ ,
\label{NVI}
\end{equation}
where $r_n$ and $V_n$ are the return and trade volume in $n$-th period, and
$\theta (x)$ is the Heavyside step-function defined as $\theta (x>0)=1$ with 
zero value otherwise. One of 
interpretations of the $PVI$ assumes that in the periods when volumes increases, 
the crowd of``uninformed" investors are in the market. These days
contribute to the $PVI$. Conversely, on days with decreasing volume, the 
"smart money" is quitely taking positions which is reflected by $NVI$.
However, it is important to remember that this is just on of possible 
interpretations of the indices and they can work because of other reasons.
We shall see in section 3 that the``classical" dynamics of the model we
analyze reveals predictions of the $NVI$ and $PVI$ indicators though the above
interpretation hardly can be used there. ``Buy" and ``sell" signals appear
when the indices cross their own moving averages. Statistical
estimations of the indicators accuracy and further references
can be found in Ref.~\cite{Achelis}.

\subsection{Efficient Market Hypothesis}
In the Introduction we formulated a conflict between TA and Efficient Market.
Indeed, if, according to EMH, any relevant information is included in prices 
already, then there is no way to predict prices from historical data, as TA 
assumes. We show now that the conflict is apparent and steams from an inaccurate
EMH definition.

Let us define  Efficient Market following~\cite{Cuthbertson} as a 
superposition of the Rational Expectation Hypothesis and Orthogonality
property. {\it Rational Expectation Hypothesis} states that:
\begin{enumerate}
\item Agents are rational, i.e. use any possibility to get more than less if the 
possibility occur.
\item There exists a perfect pricing model and all market 
participants know this model.
\item Agents have all relevant information to incorporate into the model.
\end{enumerate}
Using the model and the information the rational agents form an expectation 
value of the future return $E_t R_{t+1}$. This expectation value can differs 
from the actual value of the return $R_{t+1}$ on a estimation error
$\epsilon _t = R_{t+1} - E_t R_{t+1}$. The {\it Orthogonality property} 
implies that:
\begin{enumerate}
\item $\epsilon _{t+1}$ is a random variable which appears due to comming of
new information.
\item $\epsilon _{t+1}$ is independent on full information set $\Omega _t$ 
at time $t$ and 
$$
E_t (\epsilon _{t+1}|\Omega _t) =0 \ .
$$
\end{enumerate}
If agents have a wrong model then the model gives a systematic error and
some serial correlation of $\epsilon _{t+1}$ and $\epsilon _{\tau}$ (with
$\tau <t+1$) emerges. For example, if 
$$
\epsilon _{t+1} = \rho \epsilon _{t} + \delta_t
$$
where $\rho$ is a parameter of the serial correlation and $\delta_t$ is a white 
noise, then:
\begin{enumerate}
\item $E_t (\epsilon _{t+1}|\Omega_t ) \neq 0$ and
\item $E_t R_{t+1}$ is not a best expectation, i.e. the model is wrong.
\end{enumerate}
Indeed, we can improve the model using 
$\tilde{E}_t R_{t+1}$ as a new model expectation:
$$
\tilde{E}_t R_{t+1} = E_t R_{t+1} + \rho (R_{t} - E_{t-1} R_{t-1}) \ .
$$ 
If agents do not improve the model in this way they are irrational and there 
exists a possibility of a superprofit. That is why some tests of the market efficiency
concentrate on the existence of the serial correlations and the superprofit.
Excelent review of various tests and results can be found in~\cite{Cuthbertson}.
Almost all of them show that EMH does not hold, at least using the existing
pricing models as a candidates for the role of the perfect model.

What is important for our goal here is that the perfect pricing model 
$E_t R_{t+1}$ is formed using all relevant information and, in particular,
historical data. Let us consider an example. If arrival of new information
increased a return of a security comparing with other securities with the 
same measure of risk, then rational traders buy the profitable security
and sell others until the returns will not equalize. This equalization 
(relaxation) process is not infinitely fast, it takes some time and has 
to be accounted in the perfect pricing model. The knowledge of how the
relaxation goes can be obtained from available information and historical
data. It means that the analysis of historical data  and underlying market 
forces are extremely important to construct a model of future prices dynamics or,
more precise, the dynamics of the expected future prices~\cite{note1}.
At this point we return to the Technical Analysis, which is a set of empirical
(phenomenological) rules for expected future prices predictions and the
corresponding investment decisions. The comparison of characteristic times of
the return fluctuations and the market relaxation defines then the applicability
of TA. If we assume that the relaxation time is much smaller than the relaxation
one, we return to the simplified EMH definition with all relevant information 
included in the price and the corresponding price random walk. Following this line
we can say that it resolve the conflict between the Technical Analysis and
the Market Efficiency.

Summarizing, the Technical Analysis can be considered as a phenomenological
method of construction of a mean price model for future prices. It uses 
price history to estimate the market relaxation and is valid when 
the relaxation time is not zero. Real prices are stochastically distributed 
around the mean price and this constitutes the Efficient Market Hypothesis.
Now we show how these TA predictions for small times and
EMH realistic distribution function for prices can be obtained from the same 
model. This model is costructed in the framework of Gauge Theory of 
Arbitrage (GTA)~\cite{KI}, which we describe in the next section.

\section{GTA model}
When a mispricing appears in a market, market
speculators and arbitrageurs rectify the mistake by obtaining a profit
from it.  In the case of profitable fluctuations they move into 
profitable assets, leaving comparably less profitable ones.  This affects
prices in such a way that all assets of similar risk become equally
attractive, i.e.  the speculators restore the equilibrium.  If this
process occurs infinitely rapidly, then the market corrects the mispricing
instantly and current prices fully reflect all relevant information.
However, clearly it
is an idealization and does not hold for small enough
times~\cite{Sofianos}. Here, following~\cite{KI}, 
we give a ``microscopic" model to describe the
money flows, the equilibration and the corresponding statistical
dynamics of prices.

The general picture, sketched above, of the restoration of equilibrium in
financial markets resembles screening in electrodynamics.
Indeed, in the case of electrodynamics, negative charges move into
the region of the positive electric field, positive charges get out of the
region and thus screen the field.  Comparing this with the financial
market we can say that a local virtual arbitrage opportunity with a
positive excess return plays a role of the positive electric field,
speculators in the long position behave as negative charges, whilst the
speculators in the short position behave as positive ones. 
Movements of positive  and negative charges screen out a profitable
fluctuation and restore the equilibrium so that there is no
arbitrage opportunity any more, i.e. the speculators have eliminated the
arbitrage opportunity.

The analogy is apparently superficial, but it is not.  It was shown
in~\cite{KI} that the analogy emerges naturally in the framework of
the Gauge Theory of Arbitrage (GTA).  The theory treats a calculation of
net present values and asset buying and selling as a parallel transport
of money in some curved space, and interpret the interest rate, exchange
rates and prices of asset as proper connection components. This
structure is exactly equivalent to the geometrical structure underlying
the electrodynamics where the components of the vector-potential are
connection components responsible for the parallel transport of the
charges.  The components of the corresponding curvature tensors are the
electromagnetic field in the case of electrodynamics and the excess rate
of return in case of GTA.  The presence of uncertainty is equivalent to the
introduction of noise in the electrodynamics, i.e.  quantization of
the theory.  It allows one to map the theory of the capital market onto the
theory of quantized gauge field interacting with matter (money flow) fields.
The gauge transformations of the matter field correspond to a change of
the par value of the asset units which effect is eliminated by a gauge tuning of
the prices and rates.  Free quantum gauge field dynamics (in the absence of
money flows) is described by a geometrical random walk for the assets prices 
with the log-normal probability distribution. In general case the
consideration maps the capital market onto Quantum Electrodynamics where
the price walks are affected by money flows and resemble real trading data.

In simple terms, we consider a composite
system of price and money flows.  In this model ''money" represents
high frequency traders with a short 
characteristic trading time (investment horizon) $\Delta $ (for
the case of S\&P500 below we use 0.5 min as the smallest horizon).
The participants trade with each other and investors with longer
time horizons.  This system is characterized by the joint probability
distribution of money allocation and price.  If we neglect the
money, the price obeys the geometrical random walk which is due to
incoming information and longer time horizons traders.  The 
trader's behaviour on time step $\Delta$ at price $S$ is described by 
the decision matrix of non-normalized transition probabilities~\cite{KI}:  
\begin{equation}
\pi (\Delta) = \left( \begin{array}{cc} 1 & t_1 S^{\beta (\Delta )} \\ 
t_2 S^{-\beta (\Delta )} & 1
\end{array} \right) 
\label{P} 
\end{equation} where the upper row
corresponds to a transition to cash from cash and shares and lower row
gives corresponding probabilities for a transition to shares. Parameter $t_1$ 
and $t_2$ represent the transaction costs, bid-ask spread and can model any 
particular investor's decision making which we want to include in the model.
Below we neglect the bid-ask spread and transaction costs but model investor's 
decision as it shown below. The
parameter $\beta$ is a fitting parameter playing the role of the
effective temperature. 

At this stage different traders are independent of each
other. We introduce an interaction by making hopping elements depending 
additionally on change in traders configuration. This interaction can model, 
for example, the ``herd" behaviour for large changes and mean-reversion 
anticipation for small
changes. In this case the simplest choice of parameters $t_1$ and $t_2^{-1}$ is
\begin{equation}
t_2 = t_1^{-1} = e^{\alpha_1 (n_1/M-1/2) - \alpha_3 \delta (n_1/M -1/2)^3} \ ,
\label{t1}
\end{equation}
where $n_1$ is a number of money amounts (each amount equivalent
to the share lot) which have been left in cash
for the characteristic time $\Delta$ and $\alpha_1 ,\alpha_3$ are some 
numerical constants. It is clear that Eqn(\ref{t1}) 
describes ``herd" behaviour: for large sales (buy off) an investor is also
biased to sale (buy), i.e. to follow the ``herd". In the same time, small
$n_1/M-1/2$ says that the market is in a stable phase. Then, if some number 
of traders are selling and lowering the price, an investor considers the 
situation as a good opportunity to buy anticipating that the price finally will 
return to its stable value. We will show in section 4 that the simple model 
(\ref{P},\ref{t1}) is able to produce very accurate description of probability
distribution function of real prices~\cite{ISfin2}. 

Each trader possesses a certain lot of shares or the equivalent 
cash amount. The formulation of the model is completed by saying that 
the transition probability for the market is a product of the geometrical 
random walk weight for price and the matrices (\ref{P}) for each participant.
The total number of participants we assume equal to $M$.

The matrix $\pi (\Delta )$ has exactly the same form as the hopping matrix for
charged particles in Quantum Electrodynamics. This form can
also be derived from the assumption that traders want to maximize their
profit~\cite{KI}. In general, it is possible to introduce risk
aversion in the model but we do not do it here since it is irrelevant for
our purpose. We also do not include many investment time horizons as it was
done in~\cite{ISfin2} to get correct scaling behaviour of the probability 
distribution function. We return to this point in section 4.

There exist several models which describe pricing in market with many 
interacting agents.
The gauge model has a feature which differs it from earlier ones.
The feature is the homogeneity of the traders
set. In earlier models traders have always been divided into ``smart" 
(who trade rationally) and ``noisy" (who follow a fad)~\cite{DeLong,Bak}. We
believe that for the consideration of short times trades this
differentiation is not appropriate.  Indeed, all high-frequency market
participants are professional traders with years of experience.
Unsuccessful traders quickly leave the market and do not affect the
dynamics.  At the same time, each of the traders has their own view on the
market and their own anticipations.  That is why their particular decision
can be only modeled in a probabilistic way distributing trader's decisions 
around the rational (true) one. In this sense the traders are 
neither ``smart" (rational) nor ``noisy" but a mixture. 

Let us state the results for the model.  The constructed model allows
us to explain quantitatively the observed high-frequency return data.
In Ref~\cite{Stanley} Figs. 1,2 show the form of the distribution
function for changes in the S\&P500 market index, which is a price of the
portfolio consisting of the main 500 stocks traded on the New York
Stock Exchange. The changes in price have been normalized by the
standard deviation. In the approximation that the changes are much
smaller than the index itself, which is obeyed with very high accuracy,
the distribution function of the normalized changes can be considered as
the distribution function of the return on the portfolio, normalized by
the standard deviation of the return.  The return on the portfolio
during the period $\Delta$ is defined as $r(\Delta) = (S(t+\Delta)
-S(t))/S(t)$. In Ref~\cite{Stanley} it was also shown that 
the distribution function obeys the
scaling property and that this property is reflected in the dependence on time
of the probability to return to the origin. It was demonstrated that
for a time period between 1 min and 1000 min (two trading days) the
probability decrease as $t^{-\alpha}$ with the exponent $\alpha=0.712\pm
0.25$. Similar scaling results have been obtained in
Ref.~\cite{Breymann} for the high-frequency return for the \$/DM exchange rate
with different values of the exponent.

To get the correct scaling behaviour (see Fig 1 in Ref~\cite{ISfin2}) 
different time horizons have been introduced to the model with the same 
dynamical rules. In this part the model follows the Fractional Market 
Hypothesis (FMH)~\cite{Peters} which states that a stable market consists 
of traders with different time horizons but with identical dynamics.
The ``microscopical" electrodynamical model is a model for the dynamics.  
In the approach the FMH substitutes the information cascade suggested 
recently to explain the scaling properties of the \$/DM exchange 
rate~\cite{Breymann}. 

However, to get the realistic profile of the 
distribution function of returns it is
sufficient to use the model we describe above with the only one time horizon.
If choose $\beta $ as
$\beta =30$ and take the number of traded lots infinite, we can plot the 
probability distribution function of returns for S\&P500 as depicted on Fig.2.
The same analysis leads to similar results for the \$/DM exchange
rate~\cite{Breymann} with slightly different values of the parameters. It
is easy to see that the theoretical and observed distribution functions
coincide exactly with the observed data accuracy. {\it This demonstrates 
that the gauge model is able to produce realistic statistical description 
of real prices in Efficient Market phase}.

Now we turn to the Technical Analysis.
It is shown in Appendix that in the limit of small time $T$ the calculation of 
the transition probability is reduced to the solution of the 
following "classical" equations which define the dynamics of the system:
\begin{equation}
\begin{array}{c}
\frac{d y}{d t} = \sigma^2 \Delta \beta^2 M (1/2 -\rho) 
- 2\alpha_1\sqrt{(1-\rho )\rho} \ sinh(v + y)
+ (\frac{d (y+\alpha_1 \rho)}{d t} - \sigma^2 \beta^2 (1/2 -\rho ))(0) \ ,
\\
\frac{d v}{d t} = (\sqrt{\rho /(1-\rho)}  - \sqrt{(1-\rho) /\rho})
\ cosh(v + y)  - 2\sqrt{(1-\rho )\rho} \ sinh(v + y)\ ,
\\
\frac{d \rho}{d t} = 2\sqrt{(1-\rho )\rho} \ sinh(v + y) \ .
\end{array}
\label{class}
\end{equation}
Here $\sigma$ is a volatility of the return,
$y(t) +\alpha_1 \rho=\beta \cdot ln S(t)$ 
(it gives the return as $\frac{d (y+\alpha_1 \rho)}{d t}/\beta $),
$v(t)$ is the velocity of the money flows and
$1-\rho (t)$ is a relative number of share lots after the 
last trade. Solutions of the system (\ref{class})
are presented on Fig.2. We can see that the solutions oscillate with the time.
The period of the oscillations depends on the parameters of the model.
The return is shifted on a half of the period with respect to $\rho (r)$.
Price oscillations fade with time which leads to the market equilibration.

Our goal now is to show that the solutions of the
derived equations (\ref{class}) indeed resemble the situation 
in the real financial market and are consistent with
standard technical analysis tools. The first point here concerns 
the connection between price movements and trading volumes.

Now when we have ascertained that prices and volumes movements well reflect
the real market situation we can estimate the performance of standard TA 
indicators for our model. The first trivial example is 
the Price Rate-of-Change (ROC) indicator~\cite{Achelis} 
which is simply the return. So we get 
"buy" and "sell" signals at points where the security return line crosses some 
pre-defined levels. One can see that for our idealized description these signals 
anticipate reversals in the underlying security's price.

Two indicators we described in section 3 can be also easely recognized 
for our model. They are Positive Volume Index ($PVI$) and Negative Volume 
Index ($NVI$) given by formulae (\ref{PVI},\ref{NVI}).

Basing on the previous discussion on the volume behaviour we plot $NVI$ and
$PVI$ for our model and estimate the connection between their behaviour
and price movements (see Fig.4). First of all, one can see that both
$PVI$ and $NVI$ trend in the same direction as prices which agrees with TA 
arguments. Another important point is that 
$NVI$'s reversal point in this model always anticipate $PVI$'s ones which
is very reasonable if we
compare the behaviour of crowd-following and informed investors. Indeed, 
calculating $NVI$ we take into account only days when the volume describes
and the "smart money" is believed to take positions. $PVI$, on the contrary,
counts only days when volume increases, i.e. crowd-following investors are 
in the market. That is why it seems to be natural that $NVI$, which reflects
"smart money" behaviour, reacts early to changing of the market situation than
$PVI$. 

Previous analysis shows that {\it the gauge model is consistent with TA 
phenomenological rules of the market relaxation for small enough time}. 
To answer the question why TA toolkit works also on sufficiently large times 
we have to return to the model with many time horizons. At characteristic time
$T$ the market participants with comparable with $T$ investment horizons
are resposible for both the formation of the shape of the distribution 
function and TA predictions while only all together investment horizons produce
correct scaling behaviour.

\section{Resume}
In the paper we considered relation of the Market Efficiency and  applicability
of the Technical Analysis. We tried to show that these two issues are not
conflicting but complementary to each other. There is a practical outcome of
the conclusion. To construct statistical models of the market 
behaviour it is important to take 
into account quasideterministic market behaviour on short times. The practical 
source of information about the behaviour comes from everyday trading practice
which is summed up into technical analysis phenomenological rules. These rules
can play a role of criterium to construct effective models for the security
pricing. It means that to model a particular security it is important to 
construct the action functional in such a way that the corresponding 
"classical" equations would be in agreement with most accurate TA tools 
tested for the security.

\section*{Acknowledgments}
We are grateful to E.Klepfish and A.Stepanenko for discussions of the subject.
We thank R.Mantegna for sending us experimental data used
in Fig 1. This work was supported by the UK EPSRC Grant GR/L29156 and
the Royal Society/NATO Postdoctoral Fellowship Award.

\section*{Appendix}

In Ref~\cite{KI} it was shown that the transition probability from the state
with the price $S(0)$ and $(n_1,m_1)$ traders in (cash,shares) to the state
with the price $S(T)$ and the corresponding traders distribution $(n,m)$ 
can be written in terms of the functional integral:
\begin{equation}
\begin{array}{l}
P(S(T),(n,m)|S(0),(n_1,m_1)) = \\
\frac{1}{n!m!}\int d\psi d\bar{\psi} \ 
\bar{\psi}^{n_1}_{1,0} \bar{\psi}^{m_1}_{2,0}
{\psi}^{n}_{1,N} {\psi}^{m}_{2,N} e^{-\bar{\psi}_{N} 
{\psi}_{N} - \bar{\psi}_{0} {\psi}_{0}}\ 
I(\bar{\psi}, {\psi}, S(0), S(T))).
\end{array}
\label{pi1}
\end{equation}
The functional integral $I(\bar{\psi}, {\psi}, S(0), S(T)))$ has the form:
\begin{equation}
I(\bar{\xi}, {\xi}, S(0), S(T))) = 
\int Dy D\psi_1 D\psi_2 D\bar{\psi}_1 D\bar{\psi}_2 e^{s} \ 
\label{i1}
\end{equation}
with the action
\begin{equation}
s = -\frac{1}{2\sigma^2} \int^{T}_{0} \dot{y}^2 dt +
\int^{T}_{0} dt (\frac{d \psi^+_1}{d t} \psi_1 +
\frac{d \psi^+_2}{d t} \psi_2 + 
\frac{t_1}{\Delta} e^{\beta y} \psi_1^+ \psi_2 + \frac{t_2}{\Delta} 
e^{-\beta y} \psi_2^+ \psi_1)
\label{s1}
\end{equation}
and the boundary conditions for integration trajectories:
$$
\psi_i(0) =\xi_i \ , \quad \bar{\psi}_i =\bar{\xi}_i \ , \quad
y(0)=ln (S(0)) \ , \quad y(T)=ln (S(T)) \ .
$$
The transition amplitudes $t_{1},t_2$ are defined as (\ref{t1}):
$$
t_2 = t_1^{-1} = exp(\alpha_1 (\psi_1^+ \psi_1/M -1) 
-\alpha_3 (\psi_1^+ \psi_1/M -1)^3) \ .
$$
with $M$ is a number of traded lots (both money and shares). 
In this appendix we derive classical equations of motion (\ref{class}) 
for the market on small enough times from the functional integral 
representation for the transition probability (\ref{pi1},\ref{i1},\ref{s1}). 
To this end we first of all consider continuous limit and 
change $y$ to $\beta y - ln\  t_1$ which results in 
the following transformation of the action $s$:
$$
s = -\frac{1}{2\sigma^2 \beta^2} \int^{T}_{0} (\frac{dln\ e^y t_1}{d t})^2 dt +
\int^{T}_{0} dt (\frac{d \psi^+_1}{d t} \psi_1 +
\frac{d \psi^+_2}{d t} \psi_2 + 
\frac{1}{\Delta} e^y \psi_1^+ \psi_2 + \frac{1}{\Delta} e^{-y} \psi_1^+ \psi_1)
\ .
$$
To extract short time behaviour we now measure time $t$ in terms of smallest 
time interval in the system, i.e. in units of the time horizon $\Delta$. 
At such small times (in a stable market) the ``herd" effect cannot play
any valuable role and we can drop out the term with $\alpha_3$ in $ln\  t_1$.
Since the full number of asset units $M$ is large, 
we also change fields $\psi^+,\psi$ to the "hydrodynamical" variables
$\rho$ and $\phi$ which have meaning of density and velocity of the money flows:
$$
\psi^+_{i} = \sqrt{M \rho_i} e^{\phi_i} \ , \qquad 
\psi_{i} = \sqrt{M \rho_i} e^{-\phi_i} \ .
$$
The variable $\rho_i (t)$ is proportional to a density of the money flows 
in the point $i$ at the moment $t$, while $\phi_{1}(t) -\phi_2(t)$ gives 
the corresponding velocity of the flows. In this variables the action takes 
the form:
\begin{equation}
S(\rho ,\phi) = M \int^{T}_{0} dt (-\frac{1}{2\sigma^2 \Delta \beta^2  M} 
\dot{y + \alpha_1 \rho_1}^2 +
\frac{d \phi_1}{d t} \rho_1 +
\frac{d \phi_2}{d t} \rho_2 + 
2 \sqrt{\rho_1 \rho_2} \ cosh(\phi_1 -\phi_2 + y)) 
\label{delta}
\end{equation}
up to boundary terms which do not contribute to the equations of motion. 
The functional integral then can be 
rewritten as
$$
I(\bar{\psi}, {\psi}, S(0), S(T))) = 
\int Dy D\phi_1 D\phi_2 D \rho_1 D\rho_2 e^{s(\rho ,\phi , y)} \ .
$$
Appearance of the large external multiplier $M$ is a key point for the
calculation of the above functional integral by saddle point method. 
Indeed, if $M$ tends to infinity
the only relevant contribution to the integral are given by the "classical"
trajectories which are defined by the minimization equations:
$$
\frac{\delta s(y, \rho ,\phi)}{\delta y} = 0 \ , \qquad
\frac{\delta s(y, \rho ,\phi)}{\delta \rho_i} = 0 \ , \qquad
\frac{\delta s(y, \rho ,\phi)}{\delta \phi_i} = 0 \ .
$$
It means that the equations define the joint dynamics of prices and money flows
for short enough times. Using explicit form (\ref{delta}) it is easy to check 
that the last equations can be written as
\begin{equation}
\frac{1}{2\sigma^2 \Delta \beta^2 M}\frac{d^2 (y + \alpha_1 \rho_1)}{dt^2} + 
\sqrt{\rho_2 \rho_1} \ sinh(\phi_1 -\phi_2 + y) = 0 \ ,
\label{y}
\end{equation}
$$
-\frac{\alpha_1}{2\sigma^2 \Delta \beta^2 M}
\frac{d^2 (y + \alpha_1 \rho_1)}{dt^2}
\frac{d \phi_1}{d t} + \sqrt{\rho_2 /\rho_1} \ cosh(\phi_1 -\phi_2 + y) = 0 \ ,
$$
$$
\frac{d \phi_2}{d t} + \sqrt{\rho_1 /\rho_2} \ cosh(\phi_1 -\phi_2 + y) = 0 \ ,
$$
\begin{equation}
-\frac{d \rho_1}{d t} + 2\sqrt{\rho_2 \rho_1} \ sinh(\phi_1 -\phi_2 + y) = 0 \ ,
\frac{d \rho_2}{d t} + 2\sqrt{\rho_1 \rho_2} \ sinh(\phi_1 -\phi_2 + y) = 0 \ .
\label{rho}
\end{equation}
First important note concerns eqn.(\ref{y}). Indeed, combining eqns (\ref{rho}) and 
(\ref{y}) we find the equation 
$$
\frac{2}{\sigma^2 \Delta \beta^2 M}\frac{d^2 y}{dt^2} = 
\frac{d \rho_2}{d t} - \frac{d \rho_1}{d t} - 
\frac{2\alpha_1}{\sigma^2 \Delta \beta^2 M}\frac{d^2 \rho_1}{dt^2}
$$
which, after integration, gives us the first order differential equation
$$
\frac{d y}{d t} = \frac{M \sigma^2 \Delta \beta^2}{2} (\rho_2 -\rho_1) + 
(\frac{d (y+\alpha_1 \rho)}{d t} - 
\frac{M \sigma^2 \Delta \beta^2}{2} (\rho_2 -\rho_1))(0) \ .
$$
Second thing to note is the fact
that $\frac{d \rho_1}{d t} +
\frac{d \rho_2}{d t} =0$, i.e. $\rho_1 + \rho_2 =const \equiv 1$.
This
can be checked by taking sum of the eqns(\ref{rho}). It allows us
to express $\rho_2$ as $1-\rho_1$ and finally leads to eqns(\ref{class}):
$$
\frac{d y}{d t} = \sigma^2 \Delta \beta^2 M (1/2 -\rho) 
- 2\alpha_1\sqrt{(1-\rho )\rho} \ sinh(v + y)
+ (\frac{d (y+\alpha_1 \rho )}{d t} - \sigma^2 \beta^2 (1/2 -\rho ))(0) \ ,
$$
$$
\frac{d v}{d t} = (\sqrt{\rho /(1-\rho)}  - \sqrt{(1-\rho) /\rho})
\ cosh(v + y)  - 2\sqrt{(1-\rho )\rho} \ sinh(v + y)\ ,
$$
$$
\frac{d \rho}{d t} = 2\sqrt{(1-\rho )\rho} \ sinh(v + y) \ .
$$
Here we introduced the notations $\rho\equiv\rho_1$ 
for the relative number of ``traders" in cash and $v=\phi_2 -\phi_1$ 
for the velocity of the money flows.

\newpage

\begin{center} {Figure caption} \end{center}

\vspace{2cm}

FIG.1 (Ref~\cite{ISfin2}) Comparison of the $\Delta =1$ min 
theoretical (solid line) and
observed~\cite{Stanley} (squares) probability distribution of the return
$P(r)$.  The dashed line (long dashes) shows the gaussian distribution with the
standard deviation $\sigma$ equal to the experimental value 0.0508.
Values of the return are normalized to $\sigma$.  The dashed line (short dashes) is the
best fitted symmetrical Levy stable distribution~\cite{Stanley}.

\vspace{2cm}

FIG.2 Solution of quasi-classical equations of motion (\ref{class}). 
Solid line represents
the time dependence of $\beta ln S(t)$, dashed line shows a deviation of 
the relative number of investors 
in $\rho_1$ from its equilibrium value 0.5. Initial conditions are
$y(0)=0.5$, $v(0)=0.1$ and $\rho(0)=1/2$. Initial values of
first derivatives are equal zero. Time is measured in units of 
$\Delta$. The parameter $\alpha$ is equal to 0.5 and 
$M \sigma^2 \Delta \beta^2 =20$.

\vspace{2cm}

FIG.3 Prices and volumes from the  quasi-classical equations of motion 
(\ref{class}). Solid line represents
the time dependence of $\beta ln S(t)$, dashed line shows trading volumes
as given in the main text. 

\vspace{2cm}

FIG.4 PVI (long dashes) and NVI (short dashes) constructed from prices and 
volumes as in Fig.3 Solid line is $\beta ln S(t)$.


\begin{thebibliography}{99}
\bibitem{note1} Some authors agree that TA shall be accounted in the pricing
model but they argue that the purpose of it is to account a presence of 
``noise" traders who use TA. This scenario also leads to reemerging of
TA patterns. This does not contradict to the point of view elaborated in the
main text where the patterns are associated with collective market forces since
these forces can be caused by the quasi-rational (partially noise) traders
whose presence we account in GTA model.
\end{thebibliography}
\end{document}